\documentclass[11pt,a4paper]{article}

\usepackage[truedimen,margin=32mm]{geometry} 

\usepackage{mathrsfs}
\usepackage{amssymb}
\usepackage{amsmath}
\usepackage{ascmac}
\usepackage{amsthm}
\usepackage[pdftex]{graphicx}
\usepackage[pdftex]{color}

\usepackage{setspace}
\usepackage{natbib}
\usepackage{color}
\usepackage{titlesec}
\titleformat*{\section}{\large\bfseries}
\titleformat*{\subsection}{\it}

\def\ep{{\varepsilon}}

\def\At{\widetilde{A}}
\def\Bt{\widetilde{B}}

\def\Re{\mathbb{R}}
\def\Var{{\rm Var}}

\title{{\bf Dynamic Spatio-temporal Zero-inflated Poisson Models for Predicting Capelin Distribution in the Barents Sea }}

\date{}
\author{}

\begin{document}

\maketitle
\doublespacing

\vspace{-1.5cm}
\begin{center}
Shonosuke Sugasawa$^1$, Tomoyuki Nakagawa$^2$, Hiroko Kato Solvang$^3$, Sam Subbey$^{4,5}$ and Salah Alrabeei$^{5,4}$
\end{center}

\noindent
$^1$Center for Spatial Information Science, The University of Tokyo, Chiba, Japan\\
$^2$Department of Information Sciences, Tokyo University of Science, Chiba, Japan\\
$^3$Marine Mammals Research Group, Institute of Marine Research, Bergen, Norway\\
$^4$Research Group on Fisheries Dynamics, Institute of Marine Research, Bergen, Norway\\
$^5$Western Norway University of Applied Sciences, Bergen, Norway

\vspace{1cm}
\noindent
*Correspondence: Shonosuke Sugasawa

Center for Spatial Information Science

The University of Tokyo

5-1-5 Kashiwanoha, Kashiwa, Chiba 277-8568, Japan 

E-mail: sugasawa@csis.u-tokyo.ac.jp

\newpage
\vspace{5mm}
\begin{center}
{\bf \large Abstract}
\end{center}

We consider modeling and prediction of Capelin distribution in the Barents sea based on zero-inflated count observation data that vary continuously over a specified survey region.
The model is a mixture of two components; a one-point distribution at the origin and a Poisson distribution with spatio-temporal intensity, where both intensity and mixing proportions are modeled by some auxiliary variables and unobserved spatio-temporal effects. 
The spatio-temporal effects are modeled by a dynamic linear model combined with the predictive Gaussian process.  
We develop an efficient posterior computational algorithm for the model using a data augmentation strategy.
The performance of the proposed model is demonstrated through simulation studies, and an application to the number of Capelin caught in the Barents sea from 2014 to 2019.


\newpage
\section{Introduction}
Predicting how marine species distribution may change in response to changes in one or several exogenous factors (including from other species) in the physical environment is attractive for management and conservation. This quest has become even more timely due to recent variability in climate and the potential effect of the variability on biotic and abiotic components of the marine environment \citep[]{dalpadado2020climate,ingvaldsen2013responses}.
Scientific fisheries surveys are usually used to monitor the space-time evolution of species distributions over several years. For ecosystems with the extensive spatial extent and/or highly variable spatial oceanographic conditions (e.g., partial ice coverage), complete survey coverage may be limited or even impossible. Survey points may be sparse, and the repeated use of the same survey grid (and stations) from year to year may also be impossible. 
The combination of limited spatial coverage and sparsity of observation points means inference about how spatial distribution of species may have changed with time cannot be based exclusively on survey observations.

The data we focus on in this paper is the estimate of size (in numbers) of Barents Sea capelin at spatial points along acoustic survey transects in August-October \citep[see, e.g.,][for a description of the survey]{fall2018seasonal}. We use data in the period 2014-2019 to derive the spatial distribution in the numbers of capelin and to predict its future distribution. 
A notable feature of the dataset is that acoustic estimates are zero in many sampling locations; such characteristics and potential spatio-temporal correlations should be taken into account in the statistical modeling of the dataset to achieve stable estimates and predictions.

There are several methods for spatio-temporal modeling of zero-inflated counts.
Most of the existing methods are for area-level data \cite[e.g.][]{ver2007space,Neelon2016,torabi2017zero,ghosal2020hierarchical}, for which efficient computation algorithms are also available \citep[e.g.][]{rue2009approximate}. 
However, such techniques cannot be applied in the current situation since our samples are observed at random locations known as point-referenced data or geostatistical data \citep{banerjee2014hierarchical}.
There are some works on point-referenced spatial modeling on the standard count data \citep{diggle1998model,kelsall2002modeling}.
For point-referenced zero-inflated count data, \cite{Wang2015} proposed a Bayesian approach using Gaussian predictive processes, assuming that the sampling locations are the same over the survey period. 
Since their algorithm for posterior Bayesian inference involves rejection sampling, the computational framework may be inefficient for a large number of parameters.
On the other hand, \cite{Bradley2018} and \cite{bradley2020bayesian} developed a hierarchical spatio-temporal model using a new distribution theory.
Although the appealing feature of the method is that the fitting procedure is computationally efficient, the flexibility of the model seems limited, and it is not trivial how to account for zero-inflation.

In this work, we propose a spatio-temporal zero-inflated Poisson model, which accommodates variability in the location of capelin survey stations over time.
We consider a Bayesian estimation method that enables us to compute point estimates and uncertainty measures, such as credible intervals.   
To enhance the efficiency of the posterior computation, we use a surrogate distribution for the Poisson distribution, regarded as a reparametrized version of the negative binomial distribution. 
We then apply the Polya-gamma data augmentation \citep{Polson2013}, and most of the full conditional distributions are turned out to be of familiar forms.
The proposed modeling approach can be seen as an extension of the spatial zero-inflated negative binomial model \citep{Neelon2018}. However, unlike in \cite{Neelon2018}, our method is not characterized by an unknown over-dispersion parameter.
When the Poisson intensity includes random (spatial or time) effects, the marginal distribution of the observed response is over-dispersed. An explicit over-dispersion parameter is, therefore, not necessary in the proposed model.
Hence, we employ the negative binomial distribution as an approximated model for the Poisson distribution to develop an efficient posterior computation algorithm.  
The proposed spatio-temporal zero-inflated modeling is applied to both simulation and capelin datasets.
We will demonstrate that the proposed method can provide a more stable estimation and prediction performance than a spatio-temporal model without zero-inflation or a zero-inflated Poisson without spatio-temporal correlations.

This paper is organized as follows. 
In Section \ref{sec:model}, we introduce the proposed model and present a detailed description of the posterior computation algorithm. 
In Section \ref{sec:sim}, we check the performance of the developed posterior computation algorithm. 
In Section \ref{sec:app}, we analyze the spatio-temporal survey data of capelin distribution via the proposed spatio-temporal zero-inflated model.

\section{Spatio-temporal modeling for zero-inflated count data }
\label{sec:model}

\subsection{Model}

Suppose that count data $y_{it}$ is collected at the $i$th location in year $t$, where $i=1,\ldots, N_t$ and $t=1,\ldots,T$.
Let $x_{it}$ be a vector of covariates associated with $y_{it}$.
We also suppose that location information $s_{it}$ (typically a two-dimensional vector of longitude and latitude) is available.
Note that these settings allow that the number of samples and sampling locations could be different over $t$. 
To take account of potential zero-inflation structure of $y_{it}$, we introduce a latent binary indicator $z_{it}\in\{0,1\}$, so the distribution of $y_{it}$ is modeled as 
$$
P(y_{it}=0|z_{it}=1)=1, \ \ \ \ \  P(y_{it}=y|z_{it}=0)={\rm Po}(y;\lambda_{it}), \ \  y=0,1,\ldots,
$$ 
where ${\rm Po}(\cdot;\lambda_{it})$ is the probability mass function of the Poisson distribution with intensity $\lambda_{it}$.
We model the latent variable $z_{it}$ and intensity $\lambda_{it}$ as follows: 

\begin{equation}\label{model}
\begin{split}
&\lambda_{it}=\exp(x_{it}^{\top}\beta+u_t(s_{it})), \ \ \ \ 
z_{it}=I(x_{it}^{\top}\gamma+\xi_t(s_{it})+e_{it}>0),
\end{split}
\end{equation}
where $\beta$ and $\gamma$ are unknown regression coefficients, $u_t(s_{it})$ and $\xi_t(s_{it})$ are unobserved spatio-temporal effects and $e_{it}\sim N(0,1)$.
Note that under the model (\ref{model}), the marginal distribution of $y_{it}$ given the spatial and time effects are expressed as 
$$
f(y_{it}|\Psi)=\Phi(g_{it})\delta_0(y_{it})+\{1-\Phi(g_{it})\}{\rm Po}(y_{it};\lambda_{it}), 
$$
with $g_{it}=x_{it}^{\top}\gamma+\xi_{it}$, thereby the expectation and probability being $y_{it}=0$ (zero-count probability) are, respectively, given by 
\begin{align*}
E[y_{it}|\Psi]&=\{1-\Phi(g_{it})\}\lambda_{it}, \ \ \ \
P(y_{it}=0|\Psi)=\Phi(g_{it})+\{1-\Phi(g_{it})\}\exp(-\lambda_{it}).
\end{align*}

For modeling the unobserved spatio-temporal effects, $u_t(s)$ and $\xi_t(s)$, we first note that our framework allows different spatial location at time $t$ as in the Capelin data in Section~\ref{sec:app}.
Hence, we employ the dynamic linear model combined with Gaussian predictive process \citep{Banerjee2008} for modeling $u_t(s)$ and $\xi_t(s)$.
Let $s_1^{\ast},\ldots,s_M^{\ast}$ be a set of knots over the region, which do not change over time. 
Further, let $H(h)$ be a $M\times M$ correlation matrix with $(k,\ell)$-element being $\nu(s_k^{\ast}, s_\ell^{\ast}; h)$, where $\nu(s_1, s_2;h)$ is a valid correlation function with unknown bandwidth $h$.
In this work, we use the Gaussian correlation function: $\nu(s_1, s_2; h)=\exp\{-\|s_1-s_2\|^2/h^2\}$.
We define $D_{it}(s_{it}; h)=H(h)^{-1}V(s_{it};h)$, where $V(s;h)=(\nu(s, s_1^{\ast};h), \ldots,\nu(s, s_M^{\ast};h))\in \Re^{M}$.
Then, we model the two effects as $u_t(s_{it})=D_{it}(s_{it}; h_v)^{\top}v_t$ and $\xi_t(s_{it})=D_{it}(s_{it}; h_{\eta})^{\top}\eta_t$, where $v_t$ and $\eta_t$ are $M$-dimensional vector of temporal effects on the fixed knots.
We assume that $v_t$ and $\eta_t$ follow the multivariate random-walk processes, described as 
$$
v_t|v_{t-1}\sim N(v_{t-1}, \tau_v^{-1}H(h_{v})), \ \ \ \ \eta_t|\eta_{t-1}\sim N(\eta_{t-1}, \tau_\eta^{-1}H(h_{\eta})),
$$
for $t=1,\ldots,T$, where $\tau_v$ and $\tau_\eta$ are precision parameters, and $h_{v}$ and $h_{\eta}$ are bandwidth parameters.
Note that we set $v_0=\eta_0=0$. 
The above formulation indicates that $v_t$ and $\eta_t$ are dynamic Gaussian processes over the fixed knots, which captures both spatial and temporal correlation. 
Note that the number of the fixed knots, $M$, is an important value to control the flexibility of the predictive process and computational cost, that is, the use of small $M$ can save the computation time even under large sample sizes, but may lose flexibility for estimating spatial effects. 
The detailed specification of $M$ and setting locations of knots will be discussed later.

\subsection{Posterior computation algorithm}\label{sec:pos}
The likelihood function of the latent and model parameters is given by
\begin{align*}
&\prod_{t=1}^{T}\prod_{i=1}^{N_i}\left\{I(g_{it}>0)\delta_0(y_{it})\right\}^{z_i}\left\{I(g_{it}\leq 0){\rm Po}(y_{it};\lambda_{it})\right\}^{1-z_{it}} \exp\left\{-\frac12\left(g_{it}-x_{it}^{\top}\gamma-\xi_t(s_{it})\right)^2\right\}. 
\end{align*}
Note that the Gaussian distribution for $u(s_{it})$ and $v_t$ are not conjugate under the Poisson likelihood, so the full conditional posterior distributions of these parameters are not familiar forms, which might make the sampling steps complicated and inefficient. 
We here solve the problem by using an approximate Poisson likelihood using the negative binomial distribution, as used in \cite{hamura2021robust}.
To this end, we introduce an additional latent variable $\ep_{it}\sim {\rm Ga}(\delta,\delta)$ in the Poisson distribution, that is, $y_{it}|(z_{it}=0,\ep_{it})\sim {\rm Po}(\ep_{it}\lambda_{it})$.
Note that $E[\ep_{it}]=1$ and $\Var(\ep_{it})=1/\delta$, thereby $\ep_{it}$ degenerates at $\ep_{it}=1$ when $\delta\to\infty$. 
In other words, the augmented model reduces to the Poisson model under $\delta\to\infty$.
Then, the likelihood of $\lambda_{it}$ under the model integrating $\ep_{it}$ is given by 
\begin{equation}\label{NB}
Po(y_{it};\lambda_{it})
\approx \frac{\Gamma(y_{it}+\delta)}{\Gamma(\delta)y_i!}\frac{(\delta^{-1}\lambda_{it})^{y_{it}}}{(\delta^{-1}\lambda_{it}+1)^{y_{it}+\delta}}.
\end{equation}
We employ the above likelihood function with large $\delta>0$ as an approximated Poisson likelihood. 
The main advantage of the approximated likelihood (\ref{NB}) is that the likelihood can be further augmented by the Polya-gamma random variable \citep{Polson2013}, that is, it holds that 
$$
\frac{(\delta^{-1}\lambda_{it})^{y_{it}}}{(\delta^{-1}\lambda_{it}+1)^{y_{it}+\delta}}=2^{-(y_{it}+\delta)}e^{\kappa_{it}\psi_{it}}\int_0^{\infty}\exp\left(-\frac12\omega_{it}\psi_{it}^2\right)p_{\rm PG}(\omega_{it};y_{it}+\delta,0)d\omega_{it},
$$
where $\kappa_{it}=(y_{it}-\delta)/2$, $\psi_{it}=\log(\lambda_{it}/\delta)=x_{it}^{\top}\beta+u_t(s_{it})-\log\delta$, and $p_{\rm PG}(\cdot;b,c)$ denotes the density function of the Polya-gamma distribution.
Here $\omega_{it}$ is an additional latent parameter, and the above integral expression shows that the conditional distribution of $\psi_{it}$ given $\omega_{it}$ is Gaussian, which leads to a tractable posterior computation algorithm.

The joint posterior distribution of the latent variables and model parameters is given by 
\begin{align*}
&\prod_{t=1}^{T}\prod_{i=1}^{N_i}\left\{I(g_{it}>0)\delta_0(y_{it})\right\}^{z_i}\left\{I(g_{it}\leq 0)\exp\left(\kappa_{it}\psi_{it}-\frac12\omega_{it}\psi_{it}^2\right)p_{\rm PG}(\omega_{it};y_{it}+\delta,0)\right\}^{1-z_{it}}\\
&
\times \pi(\theta) \phi(g_{it};x_{it}^{\top}\gamma+\xi_t(s_{it}), 1)
\left\{\prod_{t=1}^{T} \phi(v_t; v_{t-1}, \tau_v^{-1}H(h_v))\phi(\eta_t; \eta_{t-1}, \tau_\eta^{-1}H(h_\eta))\right\},
\end{align*}
where $\pi(\theta)$ is a prior distribution for $\theta$.
For the prior distributions, we adopt $\beta\sim N(0, D_{\beta})$, $\gamma\sim N(0, D_{\gamma})$, $\tau_v\sim {\rm Ga}(d_{\tau_v}, d_{\tau_v})$, and $\tau_\eta\sim {\rm Ga}(d_{\tau_\eta}, d_{\tau_\eta})$ as default choices, which will be shown to be conditionally conjugate. 
We also introduce discrete uniform prior distributions for two bandwidth parameters, $h_u$ and $h_\xi$, that is, we put equal probabilities on the pre-specified candidates of knots, $\{h_1,\ldots,h_L\}$.
Under the specification, the posterior inference on unknown parameters can be conducted through posterior samples generated via the Markov Chain Monte Carlo algorithm. 
The detailed step-by-step algorithm is given as follows.

\begin{itemize}
\item[-]
{\bf Sampling of $\omega_{it}$:}
Given $z_{it}=0$, the full conditional distribution of $\omega_{it}$ is $PG(y_{it}+\delta, \psi_{it})$, where $\psi_{it}=x_{it}^{\top}\beta+u_t(s_{it})-\log\delta$.
Since the Polya-gamma distribution $PG(b,c)$ can be precisely approximated by a normal distribution under large $b$ \citep{Glynn2019} and $\delta$ is set to a large value such as $\delta=10^5$, we use $N(\tilde{b}_{it}, \tilde{c}_{it})$ as the accurate proxy of the full conditional distribution, where 
$$
\tilde{b}_{it}=\frac{y_{it}+\delta}{2\psi_{it}}\tanh\left(\frac{\psi_{it}}{2}\right),  \ \ \ \ 
\tilde{c}_{it}=\frac{y_{it}+\delta}{4\psi_{it}^3}{\rm sech}^2\left(\frac{\psi_{it}}{2}\right)\left\{\sinh(\psi_{it})-\psi_{it}\right\}.
$$

\item[-]
{\bf Sampling of $\beta$:}
The full conditional distribution of $\beta$ is $N(\At_{\beta}\Bt_{\beta}, \At_{\beta})$, where 
\begin{align*}
\At_{\beta}&=\Big(\sum_{t=1}^T\sum_{i=1}^{N_t}(1-z_{it})\omega_{it}x_{it}x_{it}^{\top}+D_{\beta}^{-1}\Big)^{-1},\\
\Bt_{\beta}&=\sum_{t=1}^T\sum_{i=1}^{N_t}(1-z_{it})x_{it}\Big\{\kappa_{it}-\omega_{it}(u_t(s_{it})-\log\delta)\Big\}.
\end{align*}

\item[-]
{\bf Sampling of $\gamma$:}
The full conditional distribution of $\gamma$ is $N(\At_{\gamma}\Bt_{\gamma}, \At_{\gamma})$, where 
\begin{align*}
\At_{\gamma}&=\Big(\sum_{t=1}^T\sum_{i=1}^{N_t}x_{it}x_{it}^{\top}+D_{\gamma}^{-1}\Big)^{-1}, \ \ \ \ 
\Bt_{\gamma}=\sum_{t=1}^T\sum_{i=1}^{N_t}x_{it}(g_{it}-\xi_t(s_{it})).
\end{align*}

\item[-]
{\bf Sampling of $v_t$ and $u_t(s_{it})$:}
For $t=1,\ldots,T-1$, the full conditional distribution of $v_t$ is $N(\At_{v_t}\Bt_{v_t}, \At_{v_t})$, where 
\begin{align*}
\At_{v_t}&=\Big\{\sum_{i=1}^{N_t}(1-z_{it})\omega_{it}D(s_{it};h_v)D(s_{it};h_v)^{\top} + 2\tau_v H(h_v)^{-1}\Big\}^{-1}, \\  
\Bt_{v_t}&=\sum_{i=1}^{N_t}(1-z_{it})D(s_{it};h_v)\Big\{\kappa_{it}-\omega_{it}(x_{it}^{\top}\beta-\log\delta)\Big\}+\tau_v H(h_v)^{-1}(v_{t-1}+v_{t+1}).
\end{align*} 
For $t=T$, $2\tau_v H(h_v)^{-1}$ in $\At_{v_t}$ is changed to $\tau_v H(h_v)^{-1}$, and the same form of $\Bt_{v_t}$ is used with $v_{T+1}=0$. 
The posterior sample of $u_t(s_{it})$ is obtained as $u(s_{it})=D(s_{it}; h_v)^{\top}v_t$.

\item[-]
{\bf Sampling of $\xi_t(s_{it})$ and $\eta_t$:}
For $t=1,\ldots,T-1$, the full conditional distribution of $\eta_t$ is $N(\At_{\eta_t}\Bt_{\eta_t}, \At_{\eta_t})$, where 
\begin{align*}
\At_{\eta_t}&=\Big\{\sum_{i=1}^{N_t}D(s_{it};h_\eta)D(s_{it};h_\eta)^{\top} + \tau_\eta H(h_\eta)^{-1}\Big\}^{-1}, \\  
\Bt_{\eta_t}&=\sum_{i=1}^{N_t}D(s_{it};h_\eta)\big(g_{it}-z_{it}^{\top}\gamma\big)+\tau_\eta H(h_\eta)^{-1}(\eta_{t-1}+\eta_{t+1}).
\end{align*} 
For $t=T$, $2\tau_\eta H(h_\eta)^{-1}$ in $\At_{\eta_t}$ is changed to $\tau_\eta H(h_\eta)^{-1}$, and the same form of $\Bt_{\eta_t}$ is used with $\eta_{T+1}=0$. 
The posterior sample of $\xi_t(s_{it})$ is obtained as $\xi_t(s_{it})=D(s_{it}; h_\eta)^{\top}\eta_t$.

\item[-]
{\bf Sampling of $h_v$:}
The full conditional distribution of $h_v$ (where $\omega_{it}$ is marginalized out) is a discrete distribution on $\{h_1,\ldots,h_L\}$ whose probability mass is proportional to
\begin{align*}
&\pi(h_v)\prod_{t=1}^T\prod_{i=1}^{N_t}\left\{\frac{(\delta^{-1}\lambda_{it})^{y_{it}}}{(\delta^{-1}\lambda_{it}+1)^{y_{it}+\delta}}\right\}^{1-z_{it}}\\
&\times |H(h_v)|^{-T/2}\exp\left\{-\frac12\tau_v\sum_{t=1}^T(v_t-v_{t-1})^\top H(h_v)^{-1}(v_t-v_{t-1})\right\},
\end{align*}
where $h_v\in\{h_1,\ldots,h_L\}$.
It is noted that $\lambda_{it}$ depends on $h_v$ through $D(s_{it}; h_{\eta})$.

\item[-]
{\bf Sampling of $h_\eta$:}
The full conditional distribution of $h_\eta$ is proportional to 
\begin{align*}
&\pi(h_{\eta})|H(h_\eta)|^{-T/2}\exp\left\{-\frac12\tau_\eta\sum_{t=1}^T (\eta_t-\eta_{t-1})^\top H(h_\eta)^{-1}(\eta_t-\eta_{t-1})
\right\}\\
& \ \ \ \ 
\times \exp\left\{-\frac12\sum_{t=1}^T\sum_{i=1}^{N_t}\left(g_{it}-x_{it}^{\top}\gamma-D(s_{it}; h_\eta)^{\top}\eta_t\right)^2\right\},
\end{align*}
where $h_\xi\in\{h_1,\ldots,h_L\}$.

\item[-]
{\bf Sampling of $\tau_v$ and $\tau_\eta$:}
The full conditional distributions of $\tau_v$ and $\tau_\eta$ are ${\rm Ga}(\delta_{\tau_v}+MT/2, \delta_{\tau_v}+\sum_{t=1}^T(v_t-v_{t-1})^\top H( h_v)^{-1}(v_t-v_{t-1})/2)$ and ${\rm Ga}(\delta_{\tau_\eta}+MT/2, \delta_{\tau_\eta}+ \sum_{t=1}^T(\eta_t-\eta_{t-1})^\top H( h_\eta)^{-1}(\eta_t-\eta_{t-1})/2))$, respectively.

\item[-]
{\bf Sampling of $g_{it}$:}
The full conditional distribution of $g_{it}$ is $N_+(x_{it}^{\top}\gamma+\xi_t(s_{it}), 1)$ if $z_{it}=1$ and $N_{-}(x_{it}^{\top}\gamma+\xi_t(s_{it}), 1)$ if $z_{it}=0$.

\item[-]
{\bf Sampling of $z_{it}$:}
The full conditional distribution of $z_{it}$ is the Bernoulli distribution with success probability $1/(1+d_{it})$, where
$$
d_{it}=\frac{\{1-\Phi(x_{it}^{\top}\gamma+\xi_t(s_{it}))\}}{\Phi(x_{it}^{\top}\gamma+\xi_t(s_{it}))\delta_0(y_{it})}\times
\frac{(\delta^{-1}\lambda_{it})^{y_{it}}}{(\delta^{-1}\lambda_{it}+1)^{y_{it}+\delta}}.
$$ 
\end{itemize}

It should be noted that all the above sampling steps are simply generated from some familiar distributions. 
Consequently, no rejection steps are required in generating posterior samples from the full conditional distributions, which would prevent high serial correlations of the posterior samples.

To complete the description of the proposed method, we specify the tuning parameters.
First, we recommend the number of knots, $M$, should be set to a moderate value to ensure low computational cost. 
For specification of the locations of knots, we suggest using the $k$-means clustering algorithm with $M$ clusters, and then set $s_1^{\ast},\ldots,s_M^{\ast}$ to be the centers of the $M$ clusters.

\subsection{Spatio-temporal prediction}\label{sec:pred}
The proposed model allows us to predict outcomes at arbitrary locations and times (e.g., next year). 
We here consider predictive distribution of $y_{0,T+1}$ at $s_{0,T+1}$.
Given the covariate vector $x_{0,T+1}$, the predictive distribution of $y_{0,T+1}$ is given by $\Phi(g_{0,T+1})\delta_0 + \{1-\Phi(g_{0,T+1})\}{\rm Po}(\lambda_{0,T+1})$, where 
\begin{align*}
&\lambda_{0,T+1}=x_{0,T+1}^\top \beta + D(s_{0,T+1};h_v)^\top v_{T+1},  \ \ \ \ 
g_{0,T+1}=x_{0,T+1}^\top \gamma + D(s_{0,T+1};h_\eta)^\top \eta_{T+1}.
\end{align*}
To generate random samples from the predictive distribution, we first generate random samples of $v_{T+1}$ and $\eta_{T+1}$ from their predictive distributions, $N(v_T, \tau_v^{-1}H(h_v))$ and $N(\eta_T, \tau_\eta^{-1}H(h_\eta))$, respectively, with posterior samples of $v_t, \eta_t, \tau_v, \tau_\eta, h_v$ and $h_\eta$. 
Then, using the posterior samples of the unknown parameters, $\beta$ and $\gamma$, we can generate random samples of $g_{0, T+1}$ and $\lambda_{0,T+1}$.

\section{Simulation study}
\label{sec:sim}

Finally, we investigate the estimation performance of the proposed method using simulated data that is generated according to the model given by (\ref{model}).
We set $T=6$ (the number of time periods) and $N_i(=N)=400$ (the number of samples in each time periods), where the setting is similar to our application in Section \ref{sec:app}.
The location information $s_{it}=(s_{it1}, s_{it2})$ was generated from the uniform distribution on $[-2,2]\times [-2,2]$.
We consider the model (\ref{model}) with a single covariate $x_{it}$ generated from $N(0, (0.5)^2)$, and the fixed regression coefficients are set as $\beta=(0.5, 0.5)$ and $\gamma=(-1.5, -1)$.   
For the two spatio-temporal effect, $u_t(s_{it})$ and $\xi_t(s_{it})$, we considered the following three scenarios: 
\begin{align*}
&{\rm (S1)}\ \ \ u_t(s_{it})=A(s_{it}; 0.5)+w_1(t), \ \ \ \ \xi_t(s_{it})=A(s_{it}; 0.9)+w_2(t), \\
&{\rm (S2)}\ \ \ u_t(s_{it})=\frac{t}{5} (0.3s_{it1}+0.3s_{it2})+\frac{t}{3}, \ \ \ \ \xi_t(s_{it})=\frac12 t(0.2s_{it1}-0.1s_{it2})+t/3, \\
&{\rm (S3)}\ \ \ u_t(s_{it})=\frac{t}{5} (0.1s_{it1}^2-0.1s_{it1}s_{it2})+\frac{t}{3}, \ \ \ \ \xi_t(s_{it})=\frac{t}{5} A(s_{it}; 0.9),
\end{align*}
where $A(s_{it}; h)$ is generated from the Gaussian processes with covariance functions given by $0.5\exp(-\|s_{it}-s_{i't'}\|^2/h^2)$, $(w_1(1),\ldots,w_1(5))=(0.4, 0.8, 1.2, 1.6, 2.0)$ and $(w_2(1),\ldots,w_2(5))=(0.5, 1, 1, 0.5, 0)$.
The spatio-temporal effects in scenario (S1) have additive forms of spatial and time effects. 
On the other hand, scenarios (S2) and (S3) have non-separable forms of spatial and time effects. 
Note that the above forms of spatio-temporal effects are not exactly the same as the ones assumed in the proposed model.

\begin{figure}[!htb]
\centering
\includegraphics[width=14cm,clip]{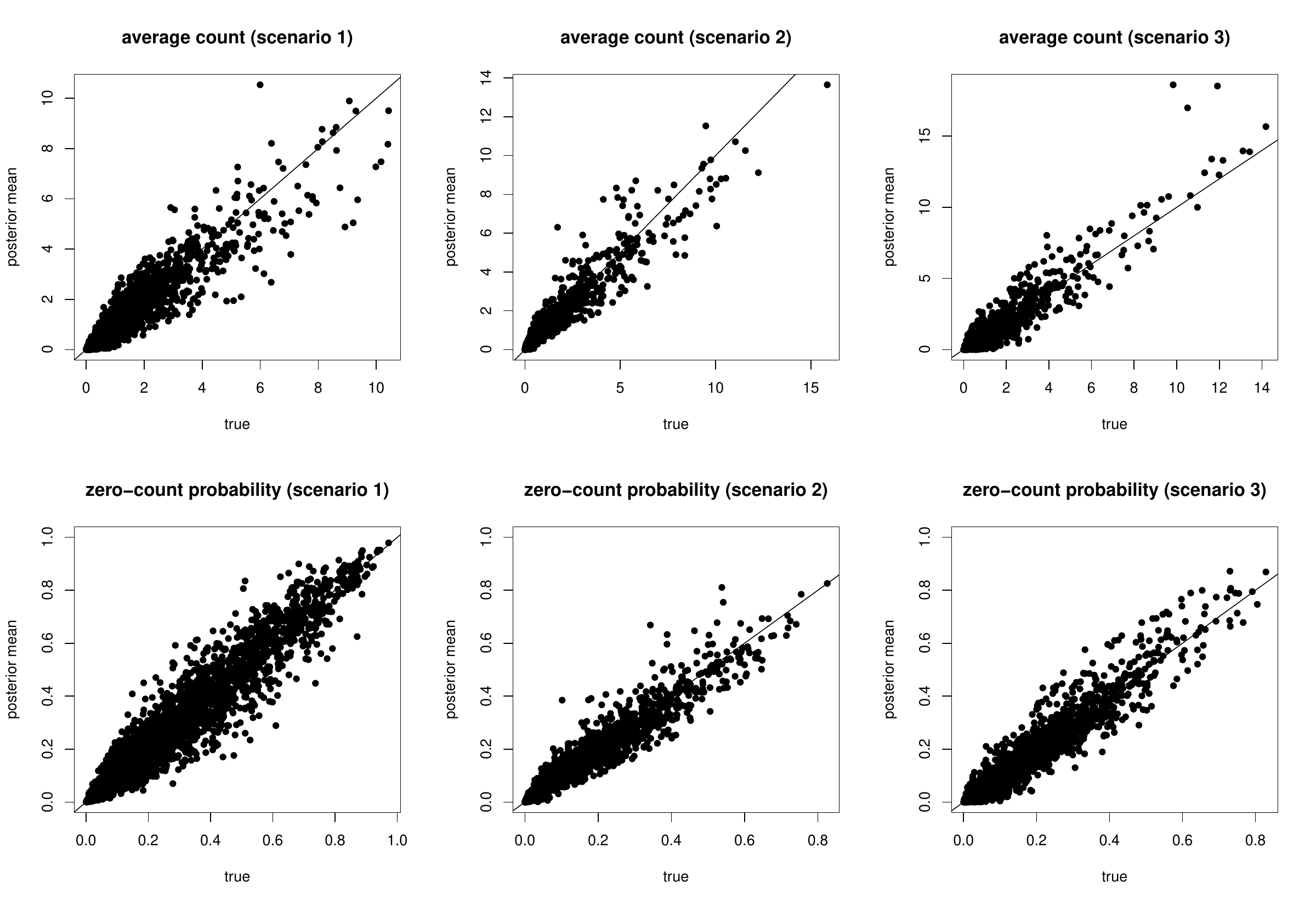}
\caption{The true values and posterior means of the average count and zero-count probability under three scenarios. }
\label{fig:sim-res}
\end{figure}

In the application of our proposed method, we set $M=100$ (the number of knots in predictive processes), $\delta=10000$ (tuning parameter for Poisson likelihood approximation), and employed default priors for the unknown parameters.
We use the Gibbs sampling procedure in Section \ref{sec:pos} to draw 40000 posterior samples after discarding the first 5000 as burn-in samples, where the convergence of our MCMC algorithm is monitored by traceplots.  
Based on the posterior samples of $\lambda_{it}$ and $g_{it}$, we computed posterior means and $95\%$ credible intervals of the average counts $E[y_{it}]$ and zero-count probability $P(y_{it}=0)$. 
In Figure~\ref{fig:sim-res}, we present scatter plots of the true values and posterior means, which show that the proposed model can precisely estimate the true average counts and zero-count probability under all three scenarios. 
In Table~\ref{tab:sim-res}, we report the empirical coverage probability (CP) and average length (AL) of the $95\%$ credible intervals and the square root of mean squared errors (RMSE) of the posterior means. 
The results show that CP values are around the nominal level in all the scenarios with reasonable AL values. 
This indicates that the proposed can flexibly capture underlying spatio-temporal structures.   
For comparison, we also applied the proposed model with $M=300$, and gave the results in Table~\ref{tab:sim-res}.
It is observed that the performance does not improve in this case, while the computation time of the model with $M=300$ is much longer than that with $M=100$. 

\begin{table}[htbp]
\caption{Coverage probability (CP) and average length (AL) of $95\%$ credible intervals and squared root of mean squared errors (RMSE) of posterior means of average count and zero-count under three scenarios.}
\label{tab:sim-res}
\begin{center}
\begin{tabular}{ccccccccccc}
\hline
&&& \multicolumn{3}{c}{average count} &&  \multicolumn{3}{c}{zero-count}\\
 &  &  & (S1) & (S2) & (S3)  &  & (S1) & (S2) & (S3)  \\
 \hline
 & CP ($\%$) &  & 94.6 & 99.5 & 97.6 &  & 96.9 & 99.2 & 98.3 \\
$M=100$ & AL &  & 1.875 & 1.358 & 1.486 &  & 0.29 & 0.182 & 0.199 \\
 & RMSE &  & 0.528 & 0.435 & 0.531 &  & 0.072 & 0.038 & 0.044 \\
 \hline
 & CP ($\%$)&  & 95.0 & 99.8 & 98.9 &  & 97.3 & 99.5 & 98.9 \\
$M=300$ & AL &  & 1.900 & 2.479 & 1.985 &  & 0.292 & 0.243 & 0.258 \\
 & RMSE &  & 0.529 & 1.084 & 0.673 &  & 0.071 & 0.050 & 0.053 \\
\hline
\end{tabular}
\end{center}
\end{table}

\section{Modeling and prediction of capelin distribution in the Barents Sea}
\label{sec:app}

\subsection{Data description}
We apply the proposed method to the Barents Sea capelin survey data in August-October in 2014-2019. The metric for determining whether capelin is matured is based on length rather than age \citep{gjosaeter2002assessment}. Based on physical and biological empirical observations, fish of at least 14cm are considered mature and capable of reproduction \citep{jourdain2021maturation}.

Our data is derived from approximately 400 sampling stations per year, with information about the number of matured/immature capelin per station. Both the locations of the sampling stations and the survey dates vary from year to year.

\begin{figure}[!htb]
\centering
\includegraphics[width=14cm,clip]{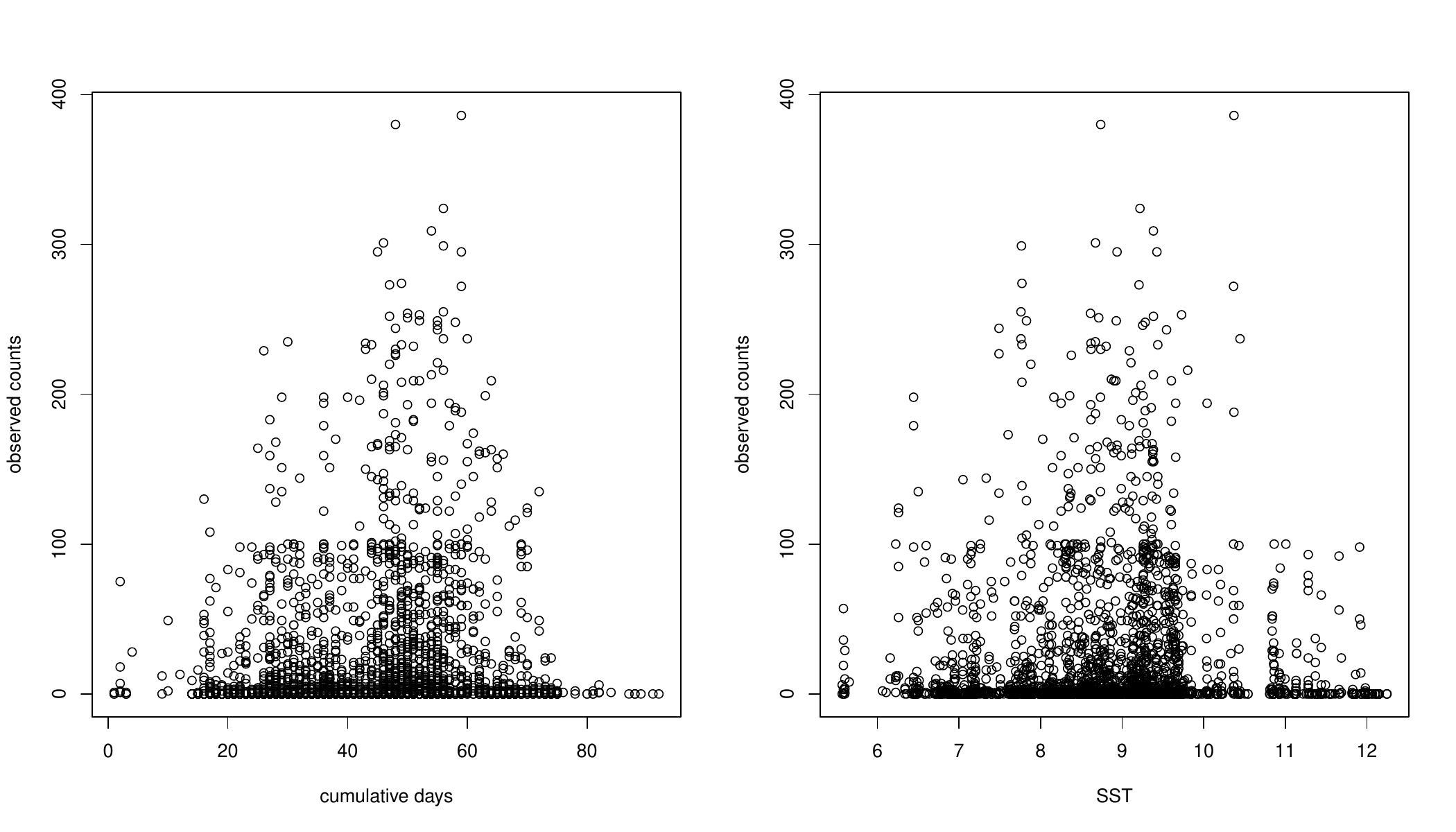}
\caption{Observed counts against cumulative days (left) and SST (right).}
\label{fig:data}
\end{figure}

As auxiliary information, we use sea surface temperature (SST) at 20m depths as the averaged values of SSTs at a depth of less than 20m at each sampling location. 
We defined 10m SST (similarly to 20m SST).
We transformed the date information to cumulative days from August 1st.  
In Figure \ref{fig:data}, we show a scatter plot of observed counts against cumulative days and 20m SST. 
It can be seen that the observed count tends to be high at $40\sim 60$ cumulative days (i.e., the sampled date is in September), and the effect of cumulative days seems nonlinear.  
Furthermore, it is observed that the observed counts and SST seem positively correlated.

\subsection{Model with nonparametric component}
Let $C_{it}$ be the cumulative day and ${\rm SST}_{it}$ be the 10m or 20m SST at the $i$th location in the $t$th year.
We consider the following model for the Poisson intensity and zero-inflation indicator: 
\begin{equation}\label{model-capelin}
\begin{split}
\log\lambda_{it}&=f_1(C_{it})+\alpha_1{\rm SST}_{it} + u_t(s_{it}), \\
z_{it}&=I\Big(f_2(C_{it})+\alpha_2{\rm SST}_{it} + \xi_t(s_{it}) + e_{it}>0\Big),
\end{split}
\end{equation} 
where $f_1(\cdot)$ and $f_2(\cdot)$ are completely unknown functions of $C_{it}$, $\alpha_1$ and $\alpha_2$ are regression coefficients of SST.
We employ the P-spline method for estimating $f_k(\cdot)  \ (k=1,2)$, that is, $f_k(\cdot)$ is estimated by the following form:
$$
f_k(x)=a_{k0}+a_{k1}x+\cdots +a_{kq}x^q + \sum_{\ell=1}^K a_{k,q+\ell}(x-\kappa_\ell)_+^q, 
$$
where $(x-c)_+=\max(x-c, 0)$, $\kappa_1,\ldots,\kappa_K$ are knots and $K$ are the number of knots.
In this analysis, we first scaled $C_{it}$ to lie on the interval $[0,1]$, and we set $q=2$, $K=9$ and $\kappa_\ell=\ell/10$. 
To avoid overfitting of the above model, we assume that $a_{k,q+\ell}\sim N(0, \tau_{Pk}^{-1})$ with unknown precision parameter $\tau_{Pk}$.

We let $\beta=(\alpha_{1},a_{10},\ldots,a_{1,q+K})^{\top}$, $\gamma=(\alpha_{2},a_{20},\ldots,a_{2,q+K})^{\top}$, and 
$$
x_{it}=({\rm SST}_{it}, 1, C_{it},\ldots,C_{it}^q, (C_{it}-\kappa_1)_+^q,\ldots,(C_{it}-\kappa_K)_+^q)^{\top}.
$$
The model (\ref{model-capelin}) can then be written in the original form (\ref{model}). 
Thus, we can employ almost the same posterior computation algorithm as in Section \ref{sec:pos}.
However, sampling steps for $\beta$ and $\gamma$ are slightly different due to the shrinkage priors for $\alpha_{k,q+\ell}$.
Under assumptions of non-informative priors for the rest of the parameters in $\beta$ and $\gamma$ (denoted by $N(0,D^{\ast}_{\beta})$ and $N(0,D^{\ast}_{\gamma})$, respectively) the sampling procedures replaces $D_{\beta}$ and $D_{\gamma}$ respectively, with ${\rm blockdiag}(D_{\beta}^{\ast}, \tau_{P1}^{-1}I_K)$ and ${\rm blockdiag}(D_{\gamma}^{\ast}, \tau_{P2}^{-1}I_K)$.
With gamma priors ${\rm Ga}(d_{\tau_{P}},d_{\tau_{P}})$ for $\tau_{Pk}$, the full conditional distribution of $\tau_{Pk}$ is given by ${\rm Ga}(d_{\tau_{P}}+K/2,d_{\tau_{P}}+\sum_{\ell=1}^Ka_{k,q+\ell}^2/2)$ for $k=1,2$, which is incorporated into the sampling algorithm in Section \ref{sec:pos}.

For model comparison, we also applied two sub-models of (\ref{model}); a spatio-temporal Poisson model (STP) without zero-inflation structure and a static zero-inflated Poisson model (ZIP) without spatio-temporal effects. 
Their MCMC algorithms can be easily obtained by slightly modifying the algorithm in Section \ref{sec:pos}.
For the three models, we generated 40000 posterior samples after discarding the first 5000 samples, where the convergence of our MCMC algorithm is monitored by traceplots.  
We computed posterior predictive loss \citep[PPL;][]{gelfand1998model} for each model, which are reported in Table \ref{tab:PPL}.
The results show that the proposed DSZIP is the best among the three models, indicating that both spatio-temporal effects and zero-inflation structures can improve overall fitting to the dataset. 
The posterior means (PM) and $95\%$ credible intervals of the regression coefficients of SST ($\alpha_1$ and $\alpha_2$) are also reported in Table \ref{tab:PPL}.
The effect ($\alpha_2$) of 10m and 20m SST on the zero-inflation probability seems to have positive effects. On the other hand, the effect ($\alpha_1$) of 10m SST on the Poisson intensity is not significant, while that of 20m SST may have a positive effect. 
The former result would be more trustworthy since the PPL with 10m SST is smaller. 
In the following discussion, we focus on the results with the 10m SST dataset.

\begin{figure}[!htb]
\centering
\includegraphics[width=14cm,clip]{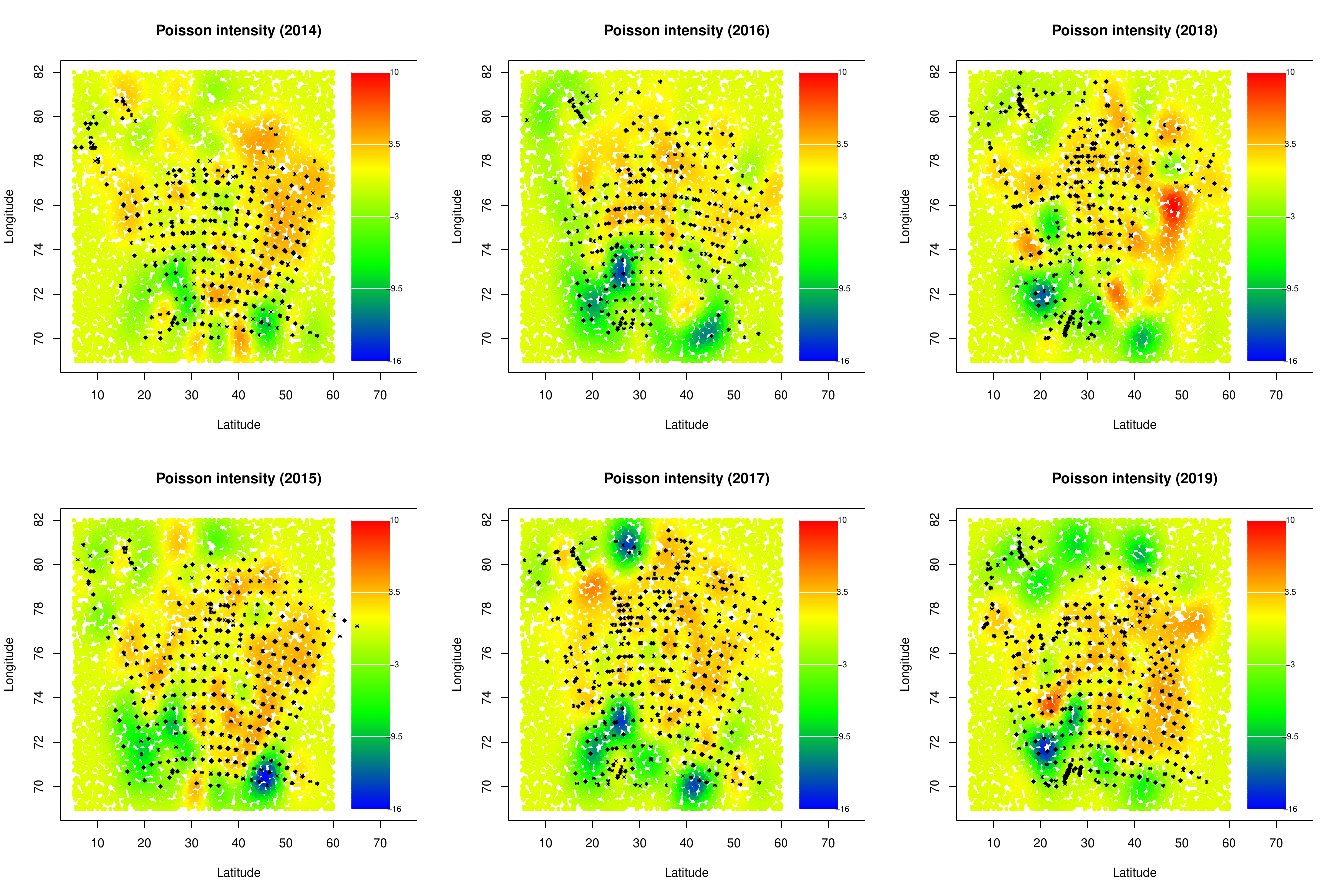}
\caption{Posterior means of the spatio-temporal effect on the Poisson intensity from 2014 to 2019. The black points indicate the sampled locations. }
\label{fig:app-ST1}
\end{figure}

\begin{figure}[!htb]
\centering
\includegraphics[width=14cm,clip]{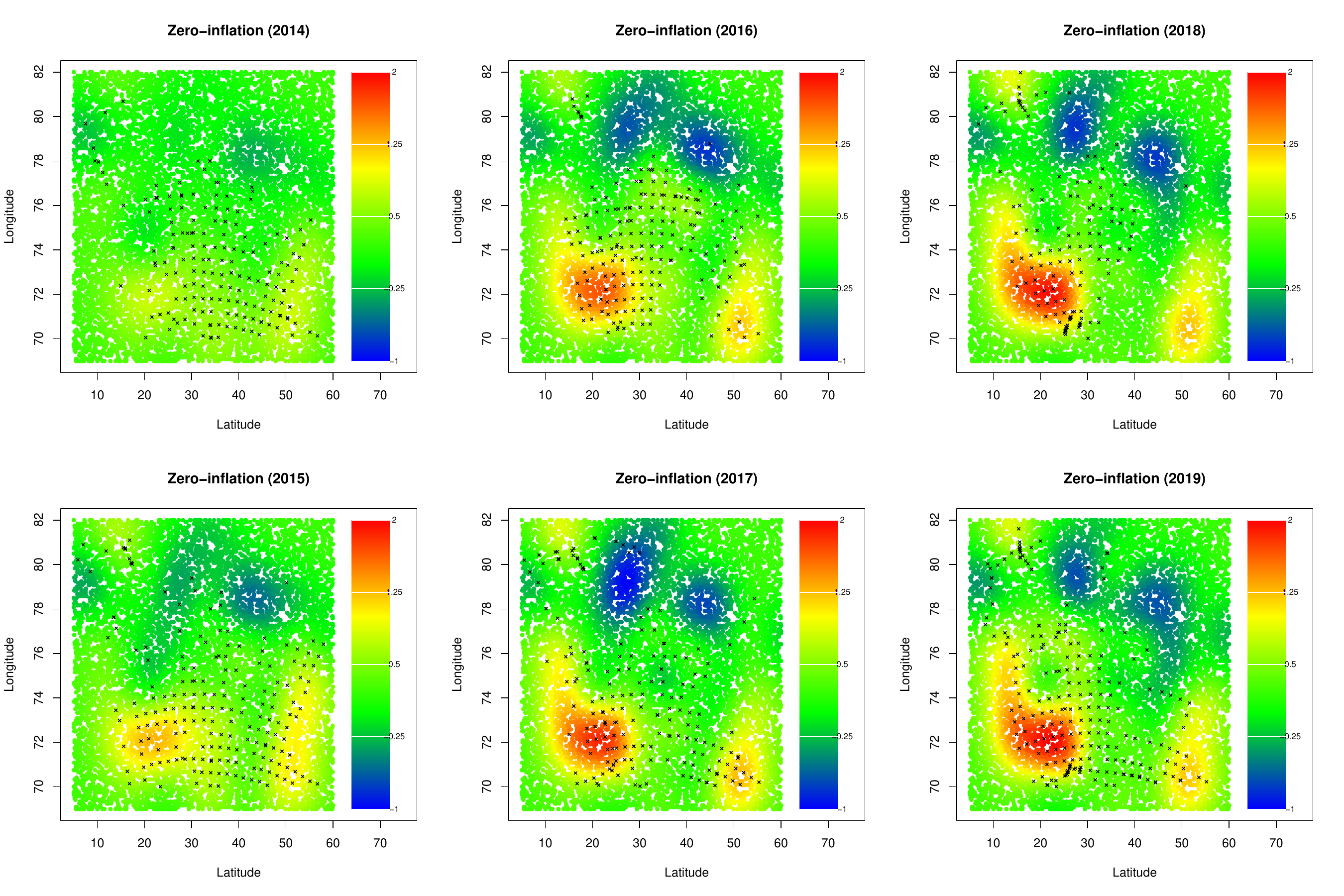}
\caption{Posterior means of the spatio-temporal effect on zero-inflation probability from 2014 to 2019. The black points indicate the sampled locations with zero-count.}
\label{fig:app-ST2}
\end{figure}

\begin{figure}[!htb]
\centering
\includegraphics[width=14cm,clip]{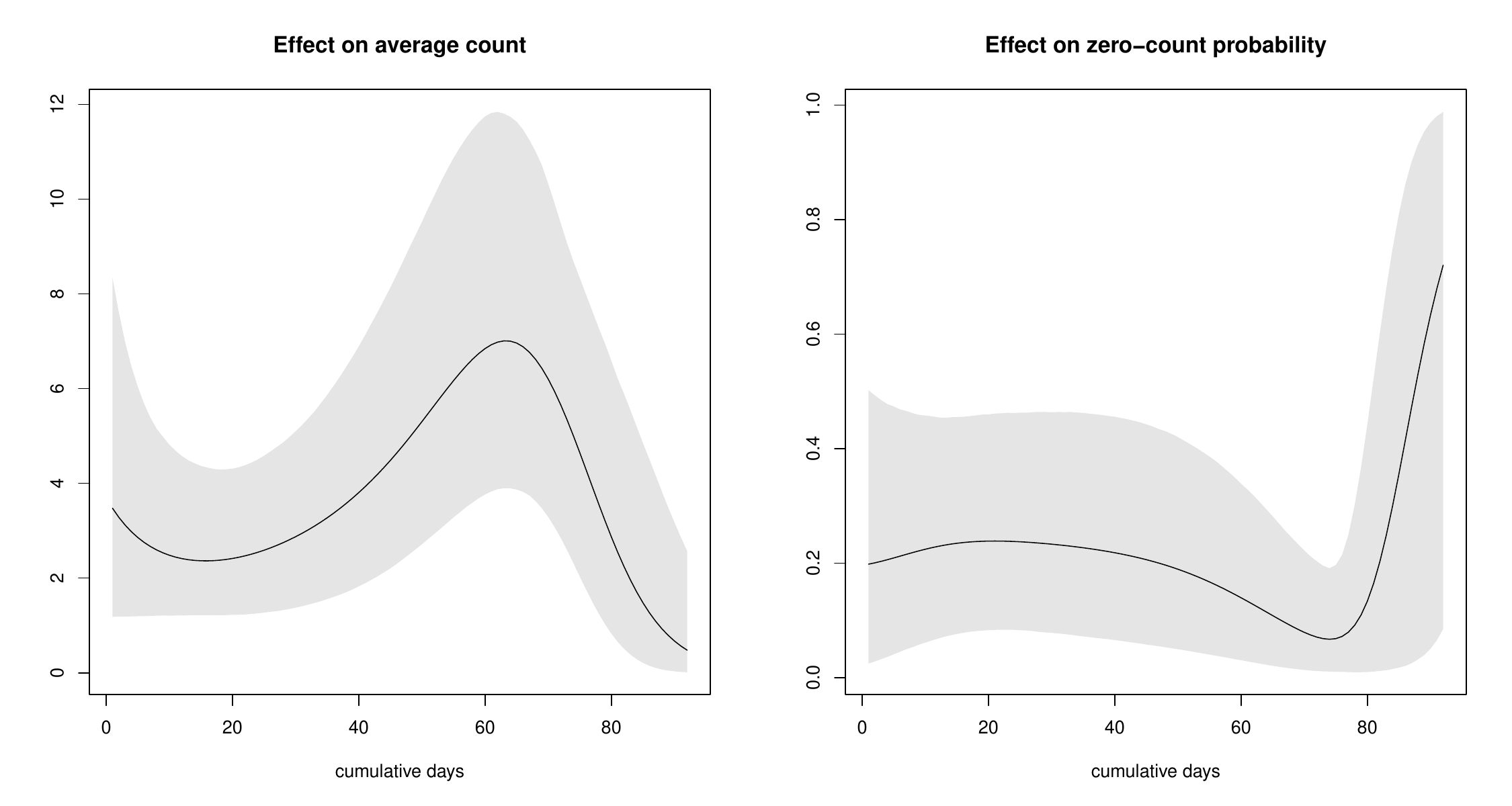}
\caption{Point-wise posterior means and $95\%$ credible intervals of the nonparametric effect of cumulative days, $f_1(\cdot)$ and $f_2(\cdot)$.}
\label{fig:app-day}
\end{figure}

The spatial plots of posterior means of the spatio-temporal effects at each year are shown in Figures~\ref{fig:app-ST1} and \ref{fig:app-ST2}.
Sampled locations are also presented in the figures.
The results indicate that the spatial effects seem to change over the years. 
We next visualize the nonparametric effect of the cumulative days. 
By fixing the spatial and time effects to 0 and setting $\alpha_1=\alpha_2=0$ in the proposed model, the expected count and zero-count probability are given by $\{1-\Phi(f_2(C))\}\exp\{f_1(C)\}$ and $\Phi(f_2(C))+\{1-\Phi(f_2(C))\}\exp(-\exp\{f_1(C)\})$, respectively, as functions of the cumulative days $C$.
The posterior means and point-wise $95\%$ credible intervals of these quantities are shown in Figure \ref{fig:app-day}.
The estimated effect in average counts seems consistent with the scatter plots of observed count given in Figure \ref{fig:data}, namely, observed counts tend to be large around $C=60$.
On the other hand, when $C>70$, the average count decreases toward 0, and zero-count probability rapidly increases, which is also consistent with Figure \ref{fig:data} in that most observed counts are 0 when $C>70$.

\begin{figure}[!htb]
\centering
\includegraphics[width=14cm,clip]{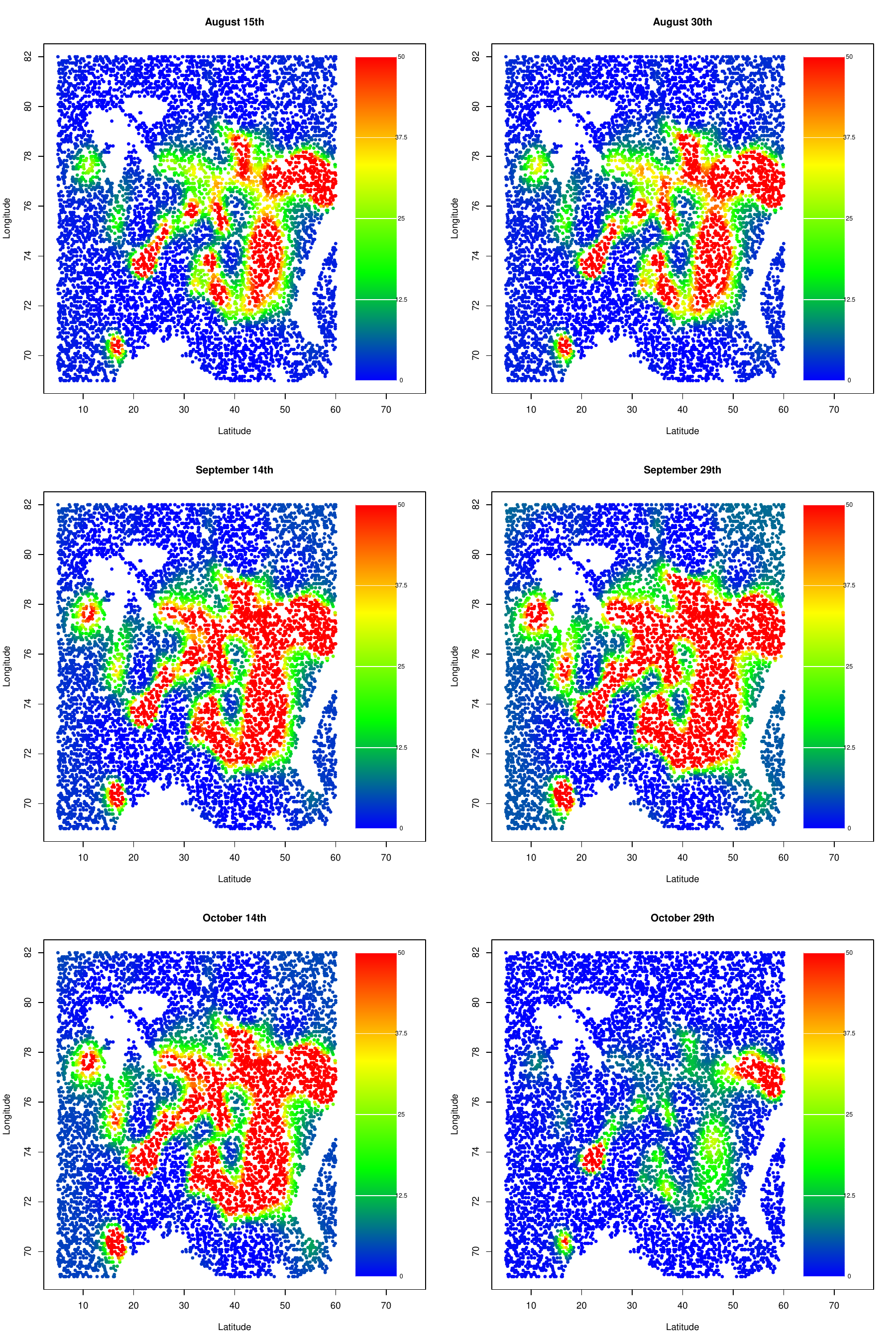}
\caption{Spatial distributions of posterior means of average counts (expected number of capelin) at 6 dates. 
Note that the average count is truncated at 50. }
\label{fig:app-intensity}
\end{figure}

\begin{figure}[!htb]
\centering
\includegraphics[width=14cm,clip]{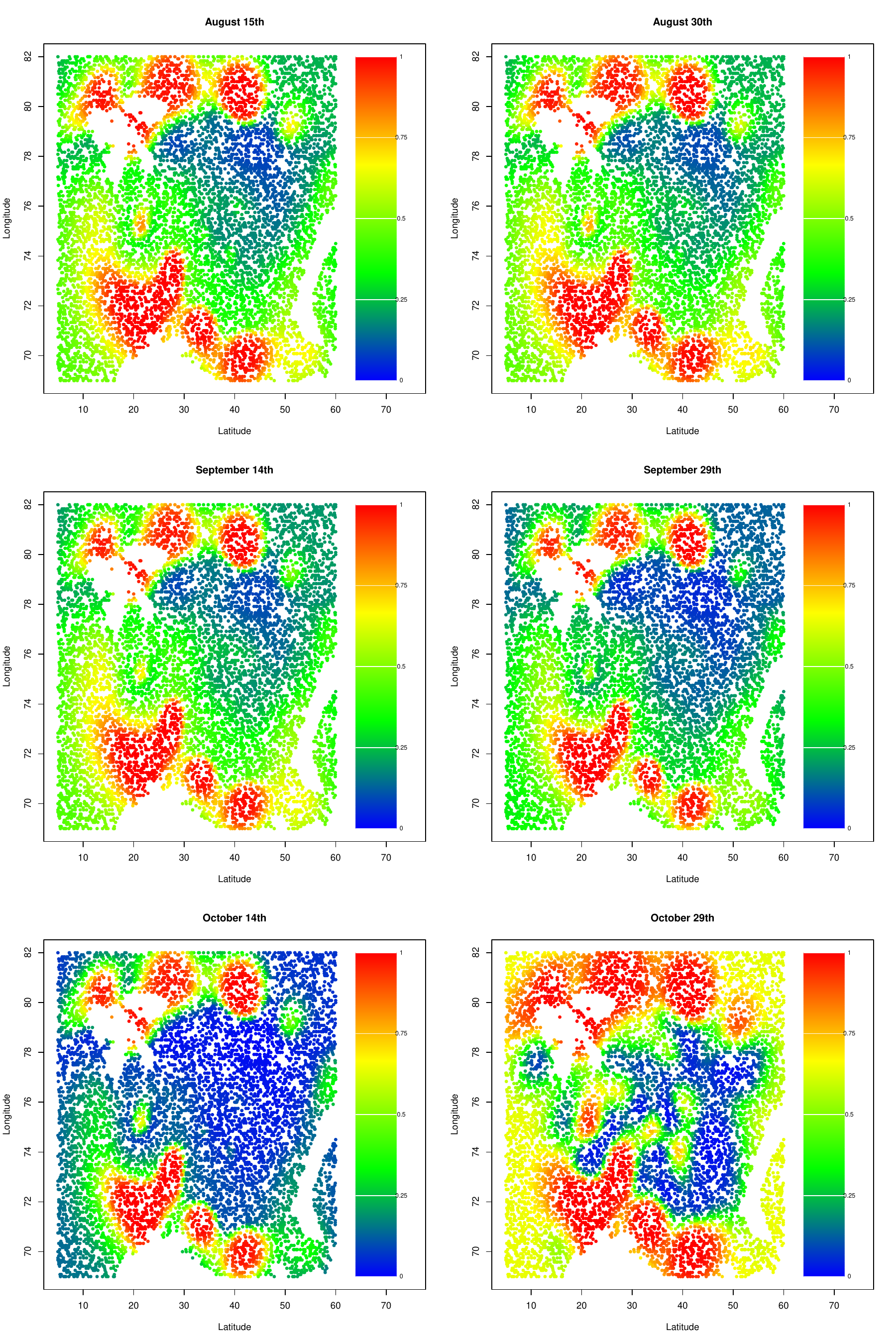}
\caption{Spatial distributions of posterior means of zero-count probability (probability that there is no capelin) at 6 dates.}
\label{fig:app-ZIP}
\end{figure}

We computed posterior means of two spatial effects in entire regions (including locations without observations) using posterior samples of $\mu_u$ and $\mu_\xi$. 
Then, using the time effect in 2019, we computed posterior means of average counts (expected number of capelin) as well as a zero-count probability (probability that there is no capelin) at 6 dates, which are presented in Figures \ref{fig:app-intensity} and \ref{fig:app-ZIP}, respectively.
Figure \ref{fig:app-intensity} indicates that some ``hot-spots" existed in the eastern part of the survey area until mid-October, which disappeared at the end of October, indicating that the average count is very small across the entire region.   
Figure \ref{fig:app-ZIP} shows that there is no capelin in the south-western region since the zero-count probability is constantly almost 1.
It is also observed that the zero-count probability is high across the entire region at the end of October, which is consistent with the results in Figure \ref{fig:app-intensity}.

\begin{table}[htbp]
\caption{Posterior mean (PM) and 95\% credible intervals of regression coefficients of and posterior predictive loss (PPL) of the three models.}
\label{tab:PPL}
\begin{center}
{\small
\begin{tabular}{ccccccccccc}
\hline
&&  \multicolumn{2}{c}{$\alpha_1$} && \multicolumn{2}{c}{$\alpha_2$}&& \multicolumn{3}{c}{PPL}\\
SST && PM & credible interval && PM & credible interval &&  DSZIP & STP & ZIP \\
\hline
10m && 0.003 & (-0.049, 0.051) && 0.073 & (-0.001, 0.145) && 1130 & 2080 & 1827 \\
20m && 0.029 & (-0.007, 0.064) && 0.048 & (-0.015, 0.111) && 1228 & 2075 & 1828 \\
\hline
\end{tabular}
}
\end{center}
\end{table}

\subsection{Spatio-temporal prediction}
Finally, we evaluated the performance of spatio-temporal prediction of the DSZIP and its two sub-models, STP and ZIP.
To this end, we left the dataset in 2019 as a validation set (or test data), and the three models with 20m SST are estimated based on the data from 2014 to 2018.
Using 20000 posterior samples after discarding the first 5000 samples, we generated random samples of counts in 2019 from their predictive distributions, along with the sampling steps in Section~\ref{sec:pred}.
We then computed the average values of the random samples to get the point prediction. 
The performance of the prediction is evaluated using the mean absolute error (MAE) and two types of mean absolute percentage error (MAPE) given by
\begin{align*}
{\rm MAE}=\frac{1}{m}\sum_{j=1}^m|y_j-\widehat{\lambda}_j|,  \ \ \ \ \ {\rm MAPE}_1=\frac{1}{m}\sum_{j=1}^m\frac{|y_j-\widehat{\lambda}_j|}{y_j+1}, \\
{\rm MAPE}_2=\Big(\sum_{j=1}^mI(y_j>0)\Big)^{-1}\sum_{j=1; y_j>0}^m\frac{|y_j-\widehat{\lambda}_j|}{y_j},
\end{align*}
where $m$ is the number of test samples, i.e. $m=499$. 
The results are reported in Table \ref{tab:prediction}, which shows that the DSZIP model provides considerably better prediction performance than the others regardless of the three different types of errors.

\begin{table}[htbp]
\caption{Three prediction errors of the three models.}
\label{tab:prediction}
\begin{center}
\begin{tabular}{ccccccccccc}
\hline
 &  & DSZIP & STP & ZIP \\
 \hline
MAE &  & 27.0 & 31.2 & 29.9 \\
MAPE$_1$ &  & 5.27 & 10.3 & 9.10 \\
MAPE$_2$ &  & 3.75 & 6.27 & 5.69 \\
\hline
\end{tabular}
\end{center}
\end{table}

\section{Concluding remarks}\label{sec:conc}
This paper has presented an algorithm for effective prediction of the spatio-temporal distribution of marine species using zero-inflated (species) count observation data and auxiliary environmental information. The algorithm has been validated with results (based on simulated and empirical data), which demonstrate the efficacy of the modeling framework.
One major significance of this paper is that it presents a computationally effective framework that facilitates studying how changes in space-time distributions may be linked to variability in environmental factors. The results in the paper are also timely, as establishing a link between spatio-temporal species dynamics and changes (e.g., temperature rise) in the physical environment is an active field of research in the marine sciences.

For modeling spatio-temporal effects, it would be possible to adopt other scalable techniques of spatial modeling, such as the nearest-neighbor Gaussian process \citep{datta2016hierarchical}. 
However, since we allow that the spatial location can be different at each time, it would not be straightforward to define a dynamic model suitable for our analysis. 
On the other hand, a class of low-rank Gaussian processes such as the predictive process is known to perform poorly in some cases \citep{stein2014limitations}, so developing a dynamic spatio-temporal model without relaying the predictive process might be valuable for future research. 
Furthermore, it may be possible to develop a more efficient algorithm by incorporating the integrated nested Laplace approximation \citep{rue2009approximate} and fast Gaussian process approximation \citep{carson2017local}.

\section*{Acknowledgement}
S. Sugasawa is supported by Japan Society for Promotion of Science (KAKENHI) grant number 18H03628, 20H00080, and 21H00699. T. Nakagawa is supported by Japan Society for Promotion of Science (KAKENHI) grant number 19K14597 and 21H00699. S. Subbey and H.K. Solvang have been supported by grants 84126 - Management Strategy for the Barents Sea, and GA19- NOR-081 from the Sasakawa foundation (Norway).

\vspace{1cm}

\bibliographystyle{chicago}
\bibliography{Ref}

\end{document}